\documentstyle[11pt,newpasp,twoside,epsf]{article}
\newcommand{\beq}{\begin{equation}}
\newcommand{\eeq}{\end{equation}}
\newcommand{\phinon}{\phi_{\rm *non}}
\newcommand{\alv}{\alpha_{\rm vir}}
\newcommand{\mvir}{M_{\rm vir}}
\newcommand{\svir}{\Sigma_{\rm vir}}

\newcommand{\msf}{m_{*f}}
\newcommand{\ecore}{\epsilon_{\rm core}}
\newcommand{\fg}{f_{\rm gas}}

\def\ion#1#2{#1$\;${\small\rm II}\relax}

\def\edcomment#1{\iffalse\marginpar{\raggedright\sl#1\/}\else\relax\fi}
\reversemarginpar

\begin{document}
\title{The Formation of Massive Stars and Star Clusters}
\author{Jonathan C. Tan}
\affil{Princeton University Observatory, Princeton NJ~08544, USA, and Dept. of Astronomy, UC Berkeley, CA 94720, USA.}
%\affil{Dept. of Astrophysical Sciences, Princeton University, Princeton NJ~08544, USA.}
%\affil{Dept. of Astronomy, UC Berkeley, CA 94720, USA.}
%\altaffiltext{*}{Present address: Dept. of Astrophysical Sciences, Princeton University, NJ~08544, USA}
\author{Christopher F. McKee}
\affil{Depts. of Physics and Astronomy, UC Berkeley, Berkeley, CA 94720, USA.}

\begin{abstract}
We model the formation of high-mass stars, specifying the accretion
rate in terms of the instantaneous and final mass of the star, the
ambient pressure of the star-forming region and the form of polytropic
pressure support of the pre-stellar gas core. The high pressures
typical of Galactic regions of massive star formation allow a
$100\:{\rm M_\odot}$ star to form in $\sim10^5\:{\rm yr}$ with a final
accretion rate $\sim 10^{-3}\:{\rm M_\odot\:yr^{-1}}$. By modeling
protostellar evolution we predict the properties of several nearby
massive protostars. We model cluster formation by applying this theory
to many stars. We use the observed intensity of outflows from
protoclusters to estimate the star formation rate, finding that
clusters take at least several free-fall times to form; for a cluster
similar to the Orion Nebula Cluster, we predict a formation timescale
$\sim 1\times 10^6\:{\rm yr}$.
\end{abstract}

\section{Introduction}

%jct further revised
%cfm revised:
If there were no massive ($m_*>8 M_\odot$) stars, the heavens would
appear quite dull. Deprived of many of the heavy elements, it is
not clear that Earth-like astronomers could even exist to contemplate
the obscurity of such a universe.
%However, such a universe would also lack many of the heavy
%elements, and it is not clear that Earth-like astronomers %--or Earth-like planets--
%could even exist to contemplate its obscurity.
%to contemplate the beauty that might have been.
%cf
Fortunately, back in our own cosmos, we can be thankful for the existence of
high-mass stars and perhaps devote some effort to understand them and
how they come about.
%jt
%jct
All
%jt
massive stars appear to form in clusters, where they are greatly
outnumbered by their lower-mass cousins, and so the questions of how
an individual massive star forms and how the cluster forms are
intimately entwined. 
%jct change in paragraph breaking

In this article
%cfm I suggest omitting the footnote
%\footnote{This article is a
%summary of the oral presentation of CFM and the poster presentation of
%JCT.} 
%cf
we first outline a theory for the formation of individual stars,
including massive ones, by accretion 
%cfm
(\S 2, taken from McKee \& Tan 2002; hereafter MT02). 
%cf
High pressures in the
star-forming region result in rapid accretion rates, which allow
massive stars to form in spite of the fierce glare of their radiation
(Wolfire \& Cassinelli 1987), and to do so swiftly before stellar
evolution has time to age the newborn 
%jct isn't ``stars'' redundant here as we have referred to ``stellar evolution''?
stars. 
%jt
In certain regions of the
Galaxy, the required high pressures are known to result from the
weight of hundreds to thousands of solar masses of strongly
self-gravitating and turbulent gas, condensed into {\it clumps} of
order a parsec across. A large fraction of this material inevitably
collapses into a stellar cluster, with most of the mass partitioned
into relatively low-mass stars.  The massive stars are expected to
form preferentially at the center of the clump, where the pressures are
highest. We apply our theory for individual star formation to model
the birth of the cluster 
%cfm
(\S\S 3 \& 4, taken from Tan \& McKee 2002a; hereafter TM02). 
%cf
Extrapolating models of outflows from
low-mass protostars to those from higher masses, we use observations
of protocluster outflow intensities to estimate the star formation
rates and cluster formation times.

\section{Collapse of Turbulent Cloud Cores and Massive Star Formation\label{S:accretion}}

Our basic assumption is that a star forms from the gravitational
collapse of a molecular cloud {\it core} that begins in approximate
hydrostatic equilibrium; in particular, it does not form by the
coalescence of low-mass stars (Bonnell, Bate, \& Zinnecker 1998) or by
the sweeping up of ambient clump material (Bonnell et al. 1997).  On
dimensional grounds we then expect that the protostellar accretion
rate is given by
\begin{equation}
\label{eq:dim}
\dot{m}_*=\phi_* \frac{m_*}{t_{\rm ff}},
\end{equation}
where $t_{\rm ff}=(3\pi/32G\rho)^{1/2}$ is the free-fall time and
$\phi_*$ is a dimensionless constant of order unity.  This equation has
the same dependence on dimensional parameters as that for isothermal
collapse, $\dot{m}_*\simeq c_{\rm th}^3/G$ (Shu 1977), if the thermal
sound speed, $c_{\rm th}$, is replaced by the virial velocity
$(Gm_*/R)^{1/2}$, since $(Gm_*/R)^{3/2}/G\propto
m_*(Gm_*/R^3)^{1/2}\propto m_*/t_{\rm ff}$.  We have shown (MT02) 
that if the collapse is spherical and self-similar, then
$\phi_*$ is quite close to unity provided that the value of $\rho$
entering $t_{\rm ff}$ is evaluated at the radius in the initial cloud
that just encloses the gas that goes into the star when its mass is
$m_*$. We summarize these results below. Collapse to the star will
naturally proceed via a disk, where we assume instabilities drive high
inflow rates if the disk mass grows 
%cfm large compared to 
to be a significant fraction of
%cf
the star's (Shu et al. 1990).

First we allow for the disruptive effect of protostellar outflows on
the gas core so that only a fraction $\epsilon_{\rm core}=m_*/M$ of
the core mass can accrete onto the star.  The value of $\epsilon_{\rm
core}$ is uncertain for massive stars, since it depends on the unknown
properties of their protostellar winds.  We assume that it is constant
over the accretion history of the star, and for numerical estimates
set $\epsilon_{\rm core}=0.5$, similar to the value
calculated for low-mass stars (Matzner \& McKee 2000).

For the core, we assume a spherical power-law ambient medium, which ensures
accretion is self-similar. With $\rho\propto r^{-k_\rho}$ and
$P\propto r^{-k_P}$, it follows that the core is a polytrope with
$P\propto \rho^{\gamma_p}$. In hydrostatic equilibrium (HSE) 
$k_\rho=2/(2-\gamma_p)$ and $k_P=\gamma_p k_\rho =
2\gamma_p/(2-\gamma_p)$ (McLaughlin \&
Pudritz 1996). Let $c\equiv (P/\rho)^{1/2}$ be the effective
sound speed.  The equation of HSE then gives $M = (k_P c^2 r)/G$ and
$\rho = A c^2/(2 \pi G r^2)$, with
$A=\gamma_p(4-3\gamma_p)/(2-\gamma_p)^2$. Equation (\ref{eq:dim}) then
implies
\begin{equation}
\label{mdot}
\dot{m}_* = \phi_* \epsilon_{\rm core} \frac{4}{\pi \sqrt{3}} k_P A^{1/2} \frac{c^3}{G},
\end{equation}
which is a generalization of the isothermal accretion rate to the
nonthermal case.

For the polytropic sphere, $\rho\propto r^{-k_\rho}$ and $M\propto
r^{3-k_\rho}$, which implies $\rho \propto M^{-k_\rho /
(3-k_\rho)}=M^{-2/(4-3\gamma_p)}$. Thus we have $\dot{m}_* \propto m_*
\rho^{1/2} \propto m_*^{1-1/(4-3\gamma_p)}$ for the mass dependence of the
accretion rate. Integration yields
\begin{equation}
m_* = m_{*f} \left(\frac{t}{t_{*f}}\right)^{4-3\gamma_p},
\label{mstar}
\end{equation}
where $m_{*f}$ is the final stellar mass, which is attained at a time
$t_{*f}$ (McLaughlin \& Pudritz 1997, hereafter MP97). Note that for
$\gamma_p<1$ the accretion rate accelerates (MP97).  Equation
(\ref{mstar}) implies
\begin{equation}
\label{mdotMP}
\dot{m}_* = (4-3\gamma_p)\frac{m_*}{t} = (4-3\gamma_p)\frac{m_{*f}}{t_{*f}}
	  \left(\frac{t}{t_{*f}}\right)^{3-3\gamma_p}.
\end{equation}
As discussed by MP97, termination of the accretion breaks the
self-similarity once the expansion wave reaches $m_{*f}$. This occurs
at a time they label $t_{\rm ew}$, which is about $0.45 t_{*f}$.
Thereafter, equation (\ref{mdotMP}) becomes approximate, but they
argue that the approximation should be reasonably good.  From
equations (\ref{eq:dim}) and (\ref{mdotMP}), the star-formation time
is
\begin{equation}
\label{tsf}
t_{*f}=\frac{(4-3\gamma_p)}{\phi_*} t_{\rm ff}.
\end{equation}
Using the results of MP97, we can evaluate $\phi_*$ in the non-magnetic
case,
\begin{equation}
\label{phi}
\phinon = (4-3\gamma_p)\frac{t_{\rm ff}}{t_{*f}} = \pi\surd 3 \left[\frac{(2-\gamma_p)^2(4-3\gamma_p)^
      {(7-6\gamma_p)/2}m_0}{8^{(5-3\gamma_p)/2}}\right]^{1/(4-3\gamma_p)},
\end{equation}
where $m_0$ is tabulated by MP97. For example, for the singular
isothermal sphere the parameter $m_0=0.975$ and $\phinon=0.975 \pi
\sqrt{3}/8=0.663$. For other values of $\gamma_p$ in the range
$0\leq\gamma_p\leq 1$, $\phinon \simeq 1.13/(1+0.7\gamma_p^2)$.  We
conclude that $\dot m_*\simeq m_*/t_{\rm ff}$ to within a factor 1.5
for spherical cores in which the effective sound speed increases
outward.

      We can estimate the effect of magnetic fields on
the accretion rate from the work of Li \& Shu (1997), who 
considered collapse of self-similar, isothermal,
magnetized, toroidal clouds.  The equilibrium surface density is
$\Sigma=(1+H_0)c_{\rm th}^2/(\pi G \varpi)$, where $\varpi$ is the cylindrical
radius and $H_0$ is a parameter that increases from zero as the
magnetic field is increased.  They show that the accretion rate is
$\dot m_*=1.05(1+H_0)c_{\rm th}^3/G$, which is 
larger than the isothermal case by about a factor $(1+H_0)$.
However, equation (\ref{eq:dim}) predicts $\dot m_*\propto
M\rho^{1/2}\propto M^{3/2}/\varpi^{3/2}\propto \Sigma^{3/2}\varpi^{3/2}
\propto (1+H_0)^{3/2}$.
To reconcile this result with the correct value,
we require $\phi_*\simeq \phinon/(1+H_0)^{1/2}$.  They regard $H_0\sim 1$ as
typical for low-mass star formation, which would lead to a reduction
in the accretion rate
%jct
(as expressed in eqs. [\ref{eq:dim}] and [\ref{mdot}])
%jt
%cfm according to the eq at the beginning of the para, \Sigma\varpi
%depends only on c_th, not on mass or density
%for a given core mass and density, 
by a factor of about 1.4.

The power-law structure of cloud cores is truncated by the ambient
pressure in the star-forming clump, which is thus equivalent to the
core surface pressure, $P_s$. We parameterize the rate of individual
core collapse in terms of this pressure, which we then relate to the
clump's surface density. A higher ambient pressure means that a core
of a given mass is truncated at higher density and thus its free-fall
and star formation timescales are shorter. 
%We have seen that the
%accretion rate also depends on the final mass of the star, $\msf$.

Since $\rho_s=P_s/c_s^2$, $M = (k_P c^2 r)/G$ and $\rho = A c^2/(2 \pi
G r^2)$, the density at the surface of the core is
\beq
\rho_s=\left(\frac{Ak_P^2\ecore^2P_s^3}{2\pi G^3\msf^2}\right)^{1/4}.
\label{rhos}
\eeq
The general expression for the accretion rate in terms of the
surface pressure and the final stellar mass can now be inferred from
equations (\ref{eq:dim}) and (\ref{rhos}):
\begin{equation}
\label{mdothighmass}
\dot{m}_* = 4.02\times10^{-4} \phi_* (A k_P^2 \ecore^2)^{\frac{1}{8}}
\left(\frac{m_{*f}}{30{\rm M_\odot}} \right)^{\frac{3}{4}} 
\left(\frac{P_s}{10^8{\rm Kcm^{-3}}}\right)^{\frac{3}{8}} 
\left(\frac{m_*}{m_{*f}}\right)^{\frac{3(2-k_\rho)}{2(3-k_\rho)}}{\rm M_\odot yr^{-1}}
\end{equation}
and the corresponding value of the star-formation time
is
\begin{equation}
\label{tsfhighmass}
t_{*f} = 7.47 \times 10^4 
\left(\frac{4-3\gamma_p}{\phi_* A^{1/8} k_P^{1/4}\ecore^{1/4}}\right)
\left(\frac{m_{*f}}{30\:{\rm M_\odot}} \right)^{1/4} 
\left(\frac{10^8\:{\rm K\:cm^{-3}}}{P_s}\right)^{3/8}\:{\rm yr}.
\end{equation}

The accretion rate and star formation time depend to some
extent on the value of $k_\rho$ in the protostellar cores.
No data are available on the structure of cores that are forming
very massive stars.  For ``high-mass'' cores in Orion, 
Caselli \& Myers (1995) find
$k_\rho\simeq 1.45$ with a dispersion of $\pm0.2$. 
According to van der Tak et al. (2000), the clumps
in which high-mass cores are embedded have values
of $k_\rho$ ranging from 1 to 2, centered around 1.5.
For numerical evaluation of the accretion rate, we shall adopt
$k_\rho=1.5$, which corresponds to $\gamma_p=2/3$, $k_P=1$, $A=3/4$,
and $\phinon=0.90$.
In this case,
the protostellar accretion rate varies linearly with time, $\dot m_*\propto t$.
So long as $A\propto
\gamma_p=2(k_\rho-1)/k_\rho$ is not near
zero, the star-formation time is relatively insensitive to $k_\rho$.
For $k_\rho=1.5$, the numerical coefficient in equation (\ref{mdothighmass}) is
$3.49\times 10^{-4}/(1+H_0)^{1/2}\:{\rm M_\odot\: yr^{-1}}$ and
$\dot{m}_* \propto (m_*/m_{*f})^{0.5}$.
Similarly, the numerical coefficient in equation (\ref{tsfhighmass})
becomes $1.72\times 10^5 (1+H_0)^{1/2}\:{\rm yr}$.  

	At present, direct observations of the total pressure at
the surface of a core that will form a massive star are not available,
so we must estimate this pressure theoretically.
Since the core is embedded in a turbulent clump, its pressure
will fluctuate, but the time-averaged surface
pressure for the typical core should be approximately the same
as the spatial average of the pressure in the clump,
$P_s=\phi_P \bar P$, where $\phi_P$ is a parameter of order unity. 
The mean pressure
in a clump is (MT02):
\beq
\bar P=\left(\frac{3\pi\fg \phi_B\alv}{20}\right)G\Sigma^2,
\label{eq:pbar}
\eeq
where $\fg$ is the fraction of the clump mass in gas 
(as opposed to stars), 
$\alv\equiv 5\langle\sigma^2\rangle R_{\rm cl}/GM_{\rm cl}\equiv \mvir/M_{\rm cl}$ is the virial
parameter, 
%cfm
$\sigma$ is the one-dimensional velocity dispersion,
$\langle x\rangle$ indicates a mass-averaged quantity,
$\phi_B=\langle c^2\rangle/\langle\sigma^2\rangle\simeq
1.3+1.5/m_A^2$ represents the contribution of the
magnetic pressure,
$m_A$ is an average of the Alfven Mach number,
and $\Sigma=M_{\rm cl}/(\pi R_{\rm cl}^2)$ is the surface density of the clump, including the
mass of any embedded stars.  In terms of $\svir$, the surface density 
determined from the virial theorem, we have $\Sigma=\svir/\alv$.

	Regions of high-mass star formation studied by Plume et
al. (1997, hereafter P97) are characterized by virial masses
$\mvir\sim 3800\:{\rm M_\odot}$ and radii $\sim 0.5\:{\rm pc}$.  The
corresponding mean column density is $\svir \simeq 1\:{\rm
g\:cm^{-2}}$; the corresponding visual extinction is $A_V=(N_{\rm
H}/2\times 10^{21}$~cm$^{-2})$~mag $=(\svir\times 214)$~mag.  These
column densities are far greater than those of giant molecular clouds (GMCs)
(0.035~g~cm$^{-2}$---Solomon et al. 1987) or of regions of low-mass star
formation (the average $\Sigma$ in the C$^{18}$O cores in Taurus
is 0.032 g cm$^{-2}$---Onishi et al. 1996).  On the other hand, the
central stellar surface density in the Orion Nebula Cluster is about 1
g cm$^{-2}$ (Hillenbrand \& Hartmann 1999).  The mean density in the
Plume et al.  clumps is $n_{\rm H}\simeq 2\times 10^5$ cm$^{-3}$.  As
discussed by MT02, the virial parameter $\alv\simeq
1.3-1.4$ for GMCs, whereas it is quite close to unity for cores that
are actively forming stars.  Although there is no direct determination
of $\alv$ for massive star-forming clumps, P97 regard the virial mass
as the most accurate, and we adopt $\alv=1$ for our numerical
estimates.

	MT02 discuss the estimates for the other
parameters, $\phi_B\simeq 2.8$, $\fg\simeq 2/3$ and $\phi_P\simeq 2$
needed to determine $P_s$.
Thus from equation (\ref{eq:pbar})
\beq
P_s=2.28\times 10^8 \left(\fg\phi_P\phi_B\alv\right)\Sigma^2~~~{\rm
	K~cm}^{-3}
	\rightarrow 8.5\times 10^8\; \Sigma^2~~~{\rm K~cm}^{-3}.
\label{eq:pbar2}
\eeq 
With $\epsilon_{\rm core}=0.5$ and $H_0=1$, the corresponding
protostellar accretion rate is
\begin{eqnarray}
%\label{mdotSigma}
\dot{m}_*&=& 4.75 \times 10^{-4} \ecore^{1/4} 
\frac{(\fg\phi_P\alv\phi_B)^{3/8}}{(1+H_0)^{1/2}}
\left(\frac{m_{*f}}{30{\rm M_\odot}}\right)^{3/4} \Sigma^{3/4} 
\left(\frac{m_*}{m_{*f}}\right)^{0.5}{\rm M_\odot yr^{-1}}\nonumber\\
&\rightarrow& 4.6\times 10^{-4}
\left(\frac{m_{*f}}{30\:{\rm M_\odot}}\right)
^{3/4} \Sigma^{3/4} 
\left(\frac{m_*}{m_{*f}}\right)^{0.5}~{\rm M_\odot\:yr^{-1}}\ \ ,
\label{mdotSigma2}
\end{eqnarray}
and the star formation time is
\begin{eqnarray}
\label{tsfSigma}
t_{*f}&=& \frac{1.26 \times 10^5}{ \ecore^{1/4}
(\fg\phi_P\alv)^{3/8}}
\left[\frac{(1+H_0)^{1/2}}{\phi_B^{3/8}}\right]
\left(\frac{m_{*f}}{30\:{\rm M_\odot}}\right)^{1/4} 
\Sigma^{-3/4}~~~ {\rm yr}\\
&\rightarrow&
1.29 \times 10^{5} 
\left(\frac{m_{*f}}{30\:{\rm M_\odot}}\right)^{1/4} 
\Sigma^{-3/4}~~~ {\rm yr}.
\end{eqnarray}
For example, with the fiducial values of
our parameters, a $100\:{\rm M_\odot}$ star forming 
in a clump with $\Sigma=1$ g cm$^{-2}$ has
a final accretion rate of $1.1\times 10^{-3}\:{\rm M_\odot\:yr^{-1}}$ and
a star-formation time of 
$1.75\times 10^5\:{\rm yr}$. 
%cfm#
Our result for the accretion rate is somewhat lower than that of
Osorio et al. (1999), who found accretion rates by matching the
infrared spectra of embedded protostars.  Our result for the 
star formation time is comparable to that of
Behrend \& Maeder
(2001), who performed an analysis based on protostellar outflows, 
although our result for the
time evolution of the accretion rate is quite different.
%cf

The above solutions apply to purely nonthermal cores. We have also
allowed for an additional thermal ($T\simeq 100\:{\rm K}$) component of
pressure support (MT02). For typical high pressure
%jct change in numerical value - applies to 100K gas
star-forming regions cores with $M\ga 1\:{\rm M_\odot}$ are dominated
by the nonthermal component. This implies that cores are
supersonically turbulent and therefore clumpy, which means we expect
the accretion rate to exhibit large fluctuations about the mean value.
%jct added note about effect of rad pressure
Radiation pressure can reduce the accretion flow to the most massive
stars from the values predicted above (Wolfire \& Cassinelli 1987; Jijina \& Adams 1996; Tan \& McKee 2002b).

\section{Protostellar Evolution, Luminosities and Outflows}

The properties of accreting protostars depend on their accretion rates
(e.g., Stahler, Shu \& Taam 1980; Stahler 1988; Palla \& Stahler 1992;
Nakano et al. 2000). In particular, the accretion luminosity $L_{\rm
acc}=f_{\rm acc} G m_*\dot{m}_*/r_*$, where $f_{\rm acc}$ is a factor
of order unity accounting for energy radiated by an accretion disk or
used to drive protostellar outflows, and the
stellar radius $r_*$ may depend sensitively on $\dot{m}_*$. Massive
stars join the main sequence during their accretion phase at a mass
that also depends on the accretion rate (Palla \& Stahler 1992). The
intensity of protostellar outflows, at least from low-mass stars, is
thought to depend on the accretion rate and the Keplerian velocity,
$v_K$, at the stellar surface (e.g. Shu et al. 1994). To predict
protostellar properties and thus compare theory with observation, we
require a model of protostellar evolution.

\begin{figure}  
\label{fig:massrad3plumeboulder}
\plotone{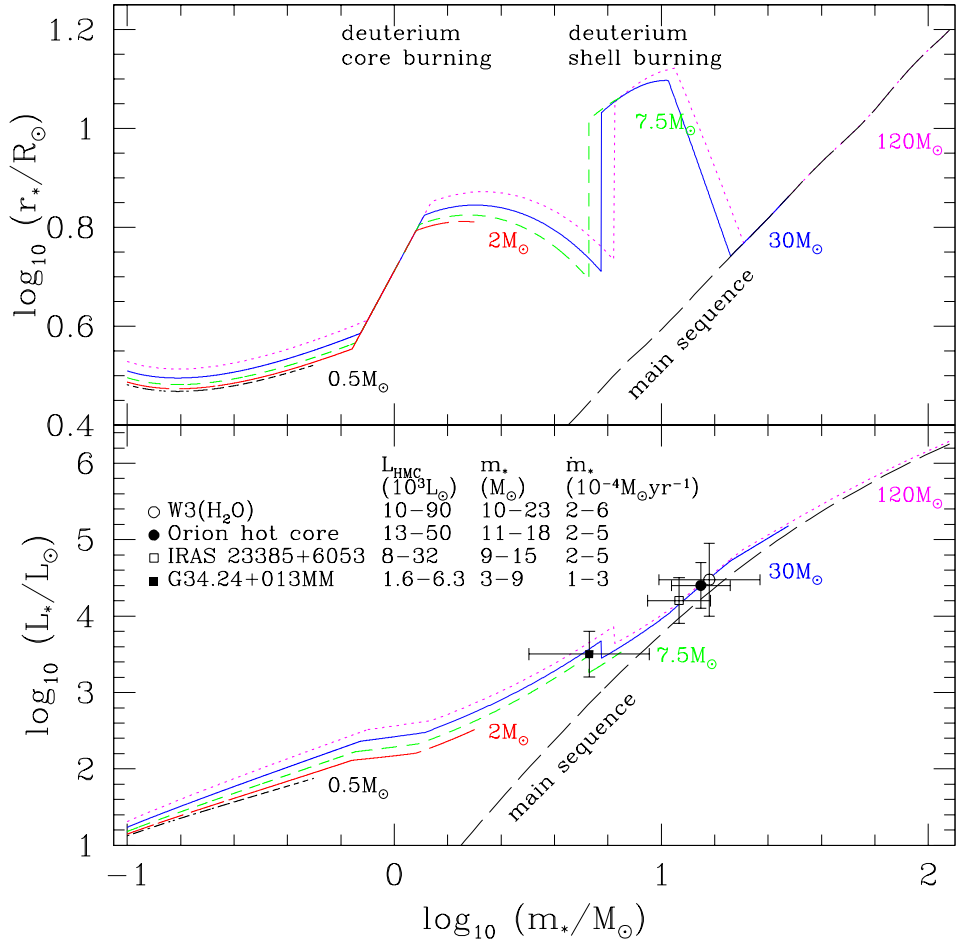}
%\plotfiddle{massrad3plumeboulder.ps}{3in}{0}{100}{100}{1}{1}
\caption{Evolution of protostellar $r_*$ and $L_*$ in a 
$\Sigma=1\:{\rm g\:cm^{-2}}$ clump. For references on HMC 
luminosities see MT02.}
\end{figure}

As previous investigations have mostly considered constant accretion
rates, we have developed a simple model for protostellar evolution
based on that of Nakano et al. (2000), which allows easy
implementation of time dependent accretion. The model accounts for the
total energy of the protostar as it accretes and dissociates matter,
expands and contracts, and, if the central temperature $T_c\ga
10^6\:{\rm K}$, burns deuterium.
%At all times the
%protostar is assumed to be a polytropic sphere of constant index. 
We have modified Nakano et al.'s model to include additional
processes, such as deuterium shell burning, and calibrated these
modifications against the more detailed calculations of Stahler (1988)
and Palla \& Stahler (1992). Protostellar evolution is followed from
$m_*=0.1\:{\rm M_\odot}$ to the end of accretion
($m_*=m_{*f}$). Massive protostars reach the main sequence (Schaller
et al. 1992) before they have finished accreting. The evolution of
protostellar radius as a star grows in mass is shown in Figure 1 for
several different final stellar masses, all forming in a clump with
$\Sigma=1\:{\rm g\:cm^{-2}}$ and $T=100\:{\rm K}$. Also shown is the
evolution in bolometric luminosities (including the contribution from
an accretion disk), and our use of these models to predict the masses
and accretion rates of protostars thought to be illuminating several
nearby hot molecular cores (HMCs).  MT02 consider the effect of
varying $k_\rho$ and $\Sigma$ on these estimates and also compare
results with Osorio et al. (1999), who have modeled the same sources.

Powerful outflows are ubiquitous from regions of both low and
high-mass star formation (e.g., Richer et al. 2000).
%; Zhang et al. 2001). 
Models of magnetically driven outflows from a disk (Blandford \& Payne
1982) and from the inner edge of a disk (Shu et al. 1994) have been
presented. A common feature of these models is the production of a
bipolar outflow with momentum distribution $p_w\propto ({\rm
sin}\:\theta)^{-2}$ for $\theta>\theta_0$, where $\theta$ is measured
from the outflow axis and $\theta_0\sim 10^{-2}$ (Matzner \& McKee
1999).
%; Shu et al. 1995). 
On scales large compared to the source (Matzner \& McKee 1999)
\begin{equation}
\label{pangular}
\frac{d \dot{p}_w}{d\Omega}= \frac{\dot{p}_w}{4\pi {\rm ln}(2/\theta_0)(1+\theta_0^2 -{\rm cos}^2\theta)}.
\end{equation}

We parameterize the total momentum flux escaping in the outflow via
\begin{equation}
\label{ew}
\dot{p}_w=f_{w,{\rm esc}} \dot{m}_w v_w = f_{w,{\rm esc}} f_w \dot{m}_* v_w = \phi_w \dot{m}_* v_K,
\end{equation}
where $f_{w,{\rm esc}}$ is the fraction of outflow momentum
that escapes from the core, $f_w=\dot{m}_w/\dot{m}_*$,
$\phi_w=f_{w,{\rm esc}} f_w v_w/v_K$ and $v_K$ is the Keplerian velocity at the
equatorial radius of the star.
%, which for rapidly rotating $n=1.5,3$
%polytropic stars is about 1.2 times the spherical radius (James 1964). 
Najita \& Shu (1994) considered the acceleration of winds from an
accreting $0.5\:{\rm M_\odot}$ protostar. For several different X-wind
model boundary conditions their results gave quite a large variation
in $f_w\approx0.1-0.8$, but only a small spread in $f_w v_w/v_K \simeq0.6$
within $\sim30$\%. Najita \& Shu (1994) also showed that $v_w$ is
approximately independent of $\theta$ so that $\dot{m}_w$ has the same
distribution as $\dot{p}_w$, and thus $f_{\rm w,esc}$ applies to wind mass
as well as momentum. If all the protostellar wind material escapes in
two cones of solid angle $\Omega_{w,{\rm esc}}$ (opening angle
$\theta_{w,{\rm esc}}$), and if core material originally inside these
cones is also swept up, then
%leads to the following relation between
%$\epsilon_{\rm core}$, $\Omega_w$ and $f_w$:
\begin{equation}
\label{epsiloncoreoutflows}
\epsilon_{\rm core}=\frac{1-2\Omega_{w,{\rm esc}}/(4\pi)}{1+f_{w,{\rm esc}} f_w}.
\end{equation}
For $\epsilon_{\rm core}=0.5$ (Matzner \& McKee 2000) and
$f_w=0.1,0.3,0.8$ we solve equations (\ref{pangular}) (with
$\dot{p}_w\rightarrow \dot{m}_w$) and (\ref{epsiloncoreoutflows}) for
$f_{w,{\rm esc}}$ and $\theta_{w,{\rm esc}}$. We find $f_{w,{\rm
esc}}=0.89, 0.86, 0.79$ and $\theta_{w,{\rm
esc}}=57^\circ,51^\circ,36^\circ$, respectively. The escape
efficiencies are quite high because of the concentration of the wind
near the outflow axis (eq. [\ref{pangular}]). Assuming the wind is
atomic, the swept-up gas is molecular and they escape together in a
momentum conserving shell, the fraction of momentum leaving the core
that is atomic in these three cases is 0.49, 0.62, 0.78, respectively.

%we have $\Omega_w=0.35\times 2\pi$, corresponding to an opening angle
%of $\theta_{\rm esc}=49.5^\circ$. 
%However, this assumes that all of
%the wind material can escape. From equation (\ref{pangular}) it is
%clear that a fraction of the mass and momentum flux is directed at
%angles $>\theta_{\rm esc}$, and this material will be
%trapped in the accretion flow and returned to the star. We solve
%equations (\ref{pangular}) and (\ref{epsiloncoreoutflows}) to find a
%self-consistent solution for $\epsilon_{\rm core}=0.5$. In this case
%$\theta_{\rm esc}=57^\circ,51^\circ,36^\circ$ for $f_w=0.1,0.3,0.8$,
%resulting in escape fractions of wind mass and momentum of 
%$f_{w,\rm esc}=0.89, 0.86, 0.79$, respectively.
%The escape efficiencies are quite
%high because of the concentration of the wind near the outflow axis
%(eq. [\ref{pangular}]). Assuming the wind is atomic, the swept-up gas
%is molecular and they escape together in a momentum conserving shell,
%the fraction of momentum leaving the core that is atomic in these
%three cases is 0.49, 0.62, 0.78, respectively.

We adopt $f_w=0.3$ and $f_{w,{\rm esc}}=0.86$ as our fiducial values
so that $\phi_w\simeq0.5$, since $f_w v_w/v_K \simeq0.6$ (Najita \&
Shu 1994). This estimate is broadly consistent with observations of
outflows from low-mass stars (Bontemps et al. 1996).
%and from protoclusters containing high-mass stars (\S\ref{S:}).
We apply the above model for low-mass protostellar outflows to stars
of all masses as they accrete. This application is uncertain: for the
X-wind model (Shu et al. 1988) to apply, the stellar surface field
strength required by a $10\:{\rm M_\odot}$ protostar with radius
$10\:{\rm R_\odot}$ accreting at a few~$\times 10^{-4}\:{\rm
M_\odot\:yr^{-1}}$ is $\sim10^3\:{\rm G}$.  The momentum
injection rates and total momenta from protostars forming in a
$\Sigma=1\:{\rm g\:cm^{-2}}$ environment are shown in Figure 2. A fit
to these results is
\begin{equation}
\label{pstarapprox}
p_w\approx p_1 m_{*f}^{\alpha_p} = 68 \left(\frac{\phi_w}{0.5}\right) \left(\frac{m_{*f}}
{\rm M_\odot}\right)^{1.4}\:{\rm M_\odot\:km\:s^{-1}}.
\end{equation}
This relation depends only weakly on $\Sigma$ (TM02).

\begin{figure}  
\label{fig:plumeoutflowboulder}
\plottwo{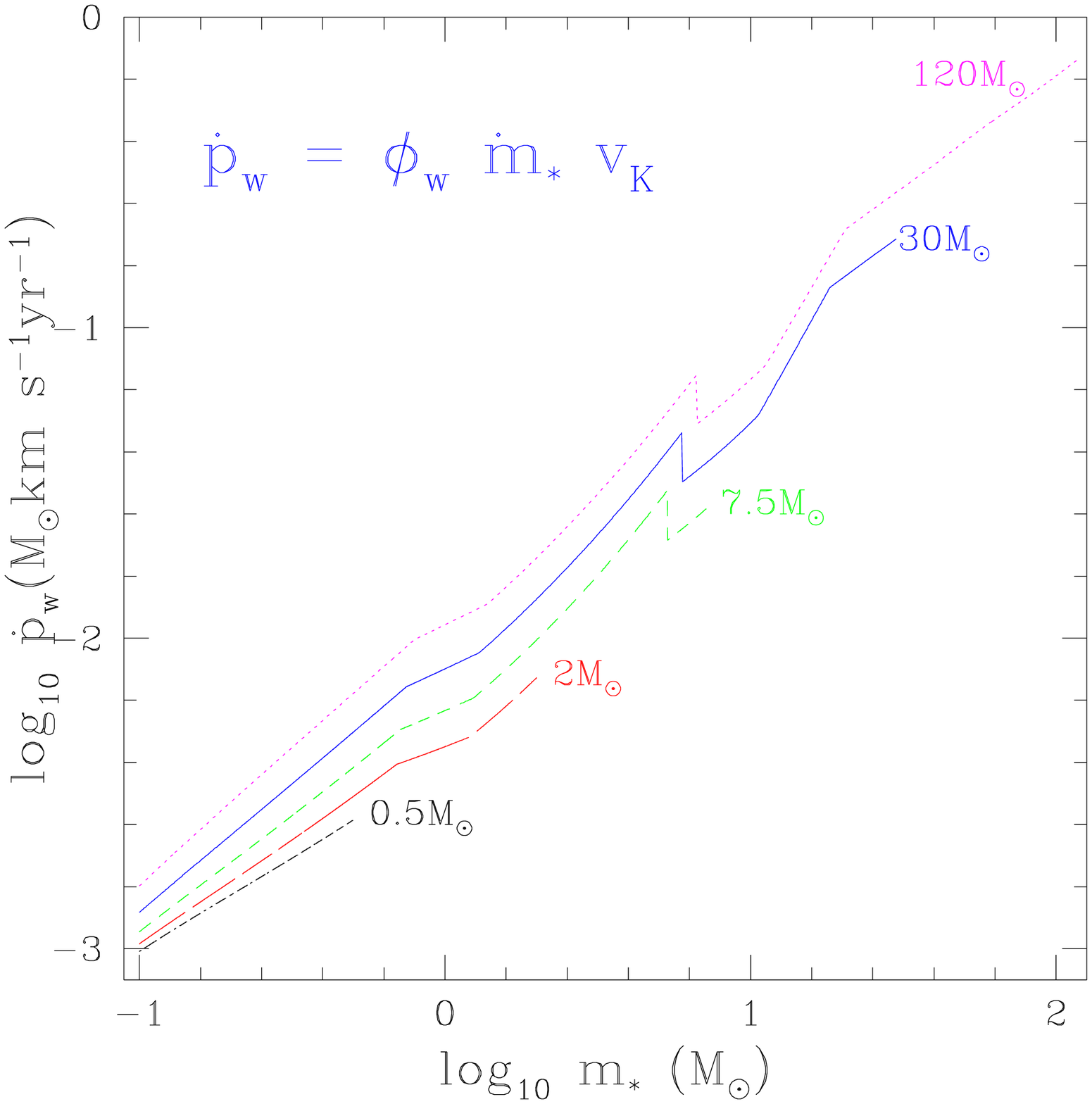}{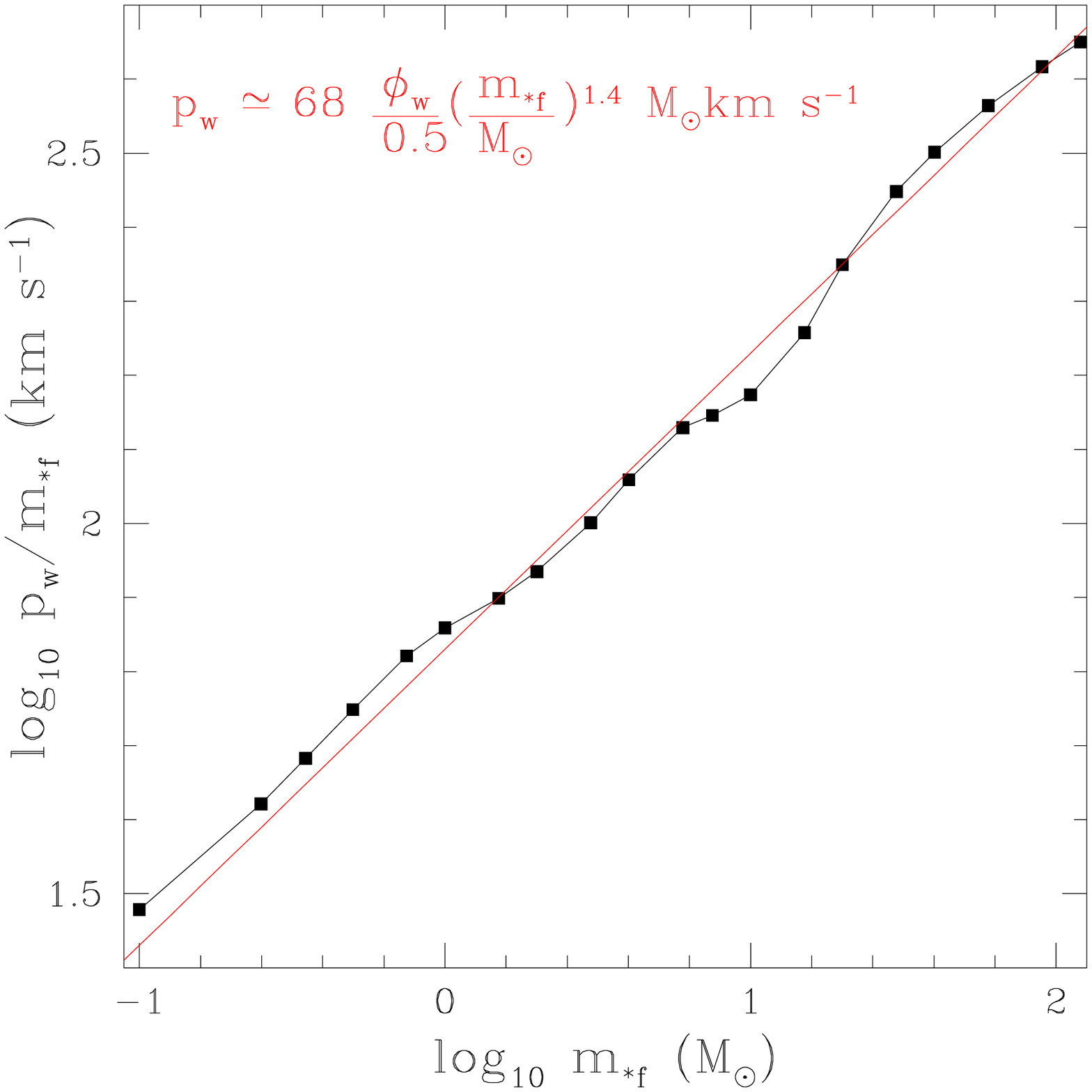}
\caption{Protostellar outflows. (a) Left: $\dot{p}_w(m_*)$ (b) Right: $p_w(m_{*f})$}
\end{figure}

\section{Probing the Rate of Star Formation in Clusters with Outflows\label{S:cluster}}

We model the formation of star clusters with ${\cal
N}_{*f}=M_{*f}/\overline{m}_{*f}$ stars, where $M_{*f}$ is the total
stellar mass and $\overline{m}_{*f}$ is the mean individual stellar mass.
Stars are drawn from a power-law initial mass function (IMF) truncated
above $m_{*u}$ and below $m_{*l}$ and of the form $dF/d\:{\rm
ln}\:m_*=A_* m_*^{-\alpha_*}$,
%cfm Not needed here:
%This may also be expressed as (McKee
%\& Williams 1997),
%\begin{equation}
%\label{IMF2}
%F(>m_*)=\frac{F_u}{\alpha_*}\left[\left(\frac{m_{*u}}{m_*}\right)^
%{\alpha_*}-1\right]%=\frac{2}{3}F_u\left[\left(\frac{m_{*u}}
%{m_*}\right)^{3/2}-1\right]\:\:(m_*\gtrsim 5)
%\end{equation}
%cf
where $F$ is the fraction of {\it all} stars that have a mass greater
than $m_*$.
%cfm
%and the normalization constant $F_u=A_*
%m_{*u}^{-\alpha_*}$.
%cf
The Salpeter (1955) IMF has 
$\alpha_*=1.35$, $A_*=0.0603$ and $\overline{m}_*=0.353\:{\rm
M_\odot}$ for $m_{*l}=0.1\:{\rm M_\odot}$ and $m_{*u}=120\:{\rm
M_\odot}$. We adopt this as our fiducial IMF. While observational
determinations of the IMF indicate deviations from this simple
power-law form at lower masses, these have little effect on most of
the feedback processes operating in a protocluster, which are
controlled by the massive stars. The total protostellar wind momentum,
$p_{\rm w,tot}$, from the forming star cluster does receive important,
approximately equal, contributions from all decades of the IMF, since
$\alpha_*\simeq\alpha_p$.  However, even in this case, the variation
caused by uncertainties in the low-mass IMF is minor. The low-mass
stars do dominate the total stellar mass, and this may be parameterized
via 
%cfm New notation
$\mu_h$, the total stellar mass per high-mass ($m_*>8\:{\rm
M_\odot}$) star, most of which end their life in a core-collapse supernova
explosion. For our adopted IMF, $\mu_h=134\:{\rm M_\odot}$, which
is close to the value for the Miller \& Scalo
(1979) IMF.
%cf

	We use Monte Carlo simulations to find the median values of the
post-accretion luminosity, $L_{*f}$, and total protostellar outflow
momentum released during formation of a cluster of mass
$M_{*f}$. Typical lower-mass clusters fail to fully sample the IMF.
For $M_{*f}/{\rm M_\odot}\ga 1600$, $L_{*f}/M_{*f}\simeq 630\:{\rm
L_\odot\:M_\odot^{-1}}$, while for $100\la M_{*f}/{\rm M_\odot}\la 1600$,
$L_{*f}/M_{*f}\simeq 436 (M_{*f}/{\rm 1000 M_\odot})^{0.8}\:{\rm
L_\odot\:M_\odot^{-1}}$.
For the outflows, with the fiducial values of the parameters
$\Sigma=1\:{\rm g\:cm^{-2}}$, $T=100\:{\rm K}$, $k_\rho=1.5$ and
$\phi_w=0.5$, we find
\begin{equation}
\label{pfit}
\frac{p_{w,\rm tot}}{M_{*f}}\simeq 87 \left(\frac{\phi_w}{0.5}\right)\left(\frac{M_{*f}}
{1000\:{\rm M_\odot}}\right)^{0.14}\:{\rm km\:s^{-1}}\:\:\:
(M_{*f}<1000\:{\rm M_\odot}).
\end{equation}
For $M_{*f}>1000\:{\rm M_\odot}$,
%cfm Note that "w" now italicized; I dropped the .1 on 87.1 here and
%in the eq above.
$p_{w,\rm tot}/M_{*f}\simeq 87(\phi_w/0.5)\:{\rm km\:s^{-1}}$. The loss of
momentum in outflow-outflow interactions would lower these estimates.
However, because of the collimation of momentum in jets
(eq.[\ref{pangular}]), we expect this effect to be relatively minor,
particularly for smaller clusters with fewer simultaneous
independent outflows.

Unless triggered by an external influence such as a supernova blast
wave, the overall rate at which stars form within a clump is not
expected to exceed $\sim M_{*f}/\overline{t}_{\rm ff}$, where
$\overline{t}_{\rm ff}=(3\pi/32G\overline{\rho})^{1/2}$ is the
free-fall time of the clump evaluated at the mean density (stars +
gas) at a typical stage in the evolution (i.e. when half the stars
have formed). We write the star formation rate 
%cfm I have inserted "typ" for typical. I have also omitted f_*, since
%it depends on the star formation efficiency but we have not discussed
%that here.
in terms of the
numerical parameter $\eta \geq 1$ as
\begin{equation}
\label{sfr1}
\dot{M}_*= \frac{M_{*f}}{\eta \overline{t}_{\rm ff}} = 
\frac{2 M_{*,{\rm typ}}}{\eta \overline{t}_{\rm ff}} = 
\frac{2 M_{\rm cl,typ}}{3\eta \overline{t}_{\rm ff}},
\end{equation}
where $M_{*,{\rm typ}}=0.5 M_{*f}$ is the typical current mass of stars,
$M_{\rm cl,typ}$ is the typical current total mass of the clump, and,
%jct
for $\epsilon_{\rm core}=0.5$, $M_{*,{\rm typ}}=M_{\rm cl,typ}/3$, if
the protostellar outflows are efficiently ejected from the clump and
most of the clump mass is in cores (see TM02).
%jt
The star formation rate is then
\begin{equation}
\label{sfr3}
\dot{M}_* = 9.7\times10^{-3} \left(\frac{1}{\eta}\right)
\left(\frac{M_{\rm cl,typ}}{1000\:{\rm M_\odot}}\right)^{3/4}
\Sigma^{3/4}\:{\rm M_\odot\:yr^{-1}}
\end{equation}
%cf
and the cluster formation time is
\begin{equation}
\label{tform}
t_{\rm cluster} = \eta \overline{t}_{\rm ff} = 6.86\times 10^4 \eta \left(\frac{M_{\rm cl,typ}}{1000\:{\rm M_\odot}}\right)^{1/4} \Sigma^{-3/4}\:{\rm yr}.
\end{equation}

We wish to estimate the cluster star formation rate, and thus $\eta$,
from observations. One method is to relate the summed total of many
individual stellar accretion events to the production of a {\it
protocluster wind}, which is the superposition of many individual
bipolar protostellar outflows. From equations (\ref{pfit}) and
(\ref{tform}) the expected momentum injection rate from a
forming star cluster is
\begin{eqnarray}
\label{pdotmedian}
\dot{p}_{w,\rm tot} & = & \frac{p_{w,\rm tot}}{t_{\rm cluster}}\nonumber\\
 & = &\cases {\frac{0.80}{\eta} \left(\frac{\phi_w}{0.5}\right) \left(\frac{M_{\rm cl,typ}}{10^3\:
{\rm M_\odot}}\right)^{0.89} \Sigma^{3/4}\:
{\rm M_\odot\: km\:s^{-1}\:yr^{-1}} 
& $\frac{M_{\rm cl,typ}}{\rm M_\odot}< 1500$,\cr
\frac{0.85}{\eta} \left(\frac{\phi_w}{0.5}\right) \left(\frac{M_{\rm cl,typ}}
{10^3\:{\rm M_\odot}}\right)^{3/4} \Sigma^{3/4}\:
{\rm M_\odot\: km\:s^{-1}\:yr^{-1}} 
& $\frac{M_{\rm cl,typ}}{\rm M_\odot}> 1500$.}
\end{eqnarray}
%cfm Note that I replace f_w,esc in the above eqs.

We compare equation (\ref{pdotmedian}) against observations of
outflows from protoclusters (Choi et al. 1993, hereafter C93). The
observational determination of $\dot{p}_{w,\rm tot}$ involves
measuring the total outflow momentum over some region and identifying
a characteristic timescale of the flow over the scale of this
region. The CO observations of C93 cover regions with sizes
similar to $R_{\rm cl}$ for which virial mass estimates have been made
from CS transitions, tracing dense gas. The spatial extent of the
outflow observations is limited by sensitivity and the flows almost
certainly continue beyond $R_{\rm cl}$. This is also evident from the
fact that the flow timescales are much shorter than $\overline{t}_{\rm
ff}$. These considerations allow us to neglect the effects of
evolution in determining an estimate of the instantaneous star
formation rate from the observed flows. Though poorly resolved, the
morphologies of the flows appear quasi-spherical, as would be expected
if there were multiple driving sources.

From CO line strengths, C93 derive the mass of the flow,
correcting for optical depth and excitation temperature effects and
assuming standard abundance ratios. The line of sight momentum is then
summed over velocity channels. We have further corrected these
estimates for projection effects (a factor of two increase for a
spherical flow). The total momentum (and mass) of the flow may be
greater by the amount contained in atomic gas. For example we have
seen in our fiducial model for protostellar winds from an individual
core that 62\% of the total momentum is in this form (i.e. resides in
gas that was processed through the accretion disk). As this flow
sweeps up gas beyond the core, the atomic fraction is reduced. 
%cfm Since you are pulling a value of Sigma/Sigma_core out of the hat,
%we might as well say that the estimate of 25% is taken from MT02.
%In lieu
%of a full treatment of this effect, we make the following simple
%estimate. We assume the gas in the clump is primarily contained in
%dense cores, with properties as described in \S\ref{S:accretion} Star
%formation is concentrated in the center of the clump, from where the
%covering factor of the cores is approximately 
%$0.75 f_{\rm gas}
%\Sigma/\Sigma_{\rm core}\simeq 0.6$. Outflows in this fraction of the
%sky have negligible atomic momentum, since the swept up gas (which is
%molecular) dominates. However the remaining 40\% of the flow retains
%its atomic momentum, which is thus $\simeq 25\%$ of the total. 
TM02 estimate that about 25\% of the total momentum is
in atomic form, so to
%cf
recover the total momentum we therefore increase the observed value by
33\%. This question may be examined observationally, though current
surveys have placed only upper limits on the fraction of atomic
momentum in the flows (Choi et al. 1994). 

	 Given estimates of mass and momentum, we determine a characteristic
flow velocity, $\overline{v}_p$, and thus an outflow timescale
$t_p=R_p/\overline{v}_p$ (here we use the data uncorrected for projection
effects, which should approximately cancel in this ratio). The value of 
$\dot{p}_{w,\rm tot}$ is
then calculated, and with knowledge of the clump's mass and surface
density, used to estimate $\eta$. The properties of our sample of protoclusters and
their outflows are listed in Tables \ref{tab:proto_cl} and
\ref{tab:proto2_cl}.

\begin{table}[h]
\caption{Properties of Protoclusters$^a$}
\label{tab:proto_cl}
\begin{center}
\begin{tabular}{cccccc}
\hline
Name & $d$ & $M_{\rm cl}$ & $R$ & $\Sigma$ & $L_{\rm bol}$\\
 & (kpc) & ($M_\odot$) & (pc) & (${\rm g\:cm^{-2}}$) & $(10^5\:{\rm L_\odot})$\\
\hline
\footnotesize S140(IRS1) & 0.9 & 128 & 0.14 & 0.43 & 0.05\\
\footnotesize GL490 & 0.9 & 91-151 & 0.06-0.083 & 1.68-1.44 & 0.017-0.024\\
\footnotesize GL2591 & 1.0 & 268-320 & 0.13-0.11 & 1.05-1.76 & 0.225\\
\footnotesize NGC2071 & 0.39 & 484 & 0.69 & 0.068 & 0.0052\\
\footnotesize W3(Main)$^b$ & 2.3 & 610 & 0.35 & 0.33 & 5.2\\
\footnotesize Cep A & 0.73 & 787 & 0.14 & 2.67 & 0.17\\
\footnotesize W28(A2) & 2.0 & 982 & 0.18 & 2.0 & 1.3-1.9\\
\footnotesize NGC6334 I(N) & 1.7 & $\sim$1700 & 0.37 & 0.83 & $<0.1$\\
%W75N & 2 & 933 & 0.17 & 2.15 & 1.4 & & & & 0.4176
\hline
\end{tabular}
\end{center}
\footnotesize $^a$ Data and lists of primary references for all
sources except NGC6334 I(N) (Megeath \& Tieftrunk 1999) are in C93,
P97, and van der Tak et al. (2000).\\ \footnotesize $^b$ W3(Main)
appears to be a relatively evolved system, containing numerous
H II regions, so the virial mass estimates based on observations
(P97) of the gas today probably underestimate the initial mass. This
can explain why 
%cfm: Note replacement of f_* by 1/3:
$M_{\rm cl}/3$ is an insufficient stellar mass to
account for $L_{\rm bol}$.
%\footnotesize $^b$ Evaluated only for the lower-mass estimate.\\
%\footnotesize $^c$ Early evolution: no atomic momentum flux correction made in calculating $\eta$.\\
%\footnotesize $^d$ Thought to be in an extremely early stage of evolution and so not included in average.\\
\end{table}

\begin{table}[h]
\caption{Observed Outflows from Protoclusters$^a$}
\label{tab:proto2_cl}
\begin{center}
\begin{tabular}{cccccc}
\hline
Name & $R_p$ & $t_p$ & $p_{w, \rm tot}$ & $\dot{p}_{w,\rm tot}$\\
 & \footnotesize HV(EHV)$^b$ & \footnotesize HV(EHV) &\footnotesize HV(EHV) &\footnotesize HV(EHV)[TOT]$^c$ & $\eta$\\
 &\scriptsize (pc) &\scriptsize ($10^3$~yr) &\scriptsize $({\rm M_\odot km\:s^{-1}})$ &\scriptsize $(10^{-3}{\rm M_\odot km\:s^{-1}yr^{-1}})$ & \\
\hline
\footnotesize S140(IRS1) &\footnotesize 0.20 (0.15) &\footnotesize 31 (6.6) &\footnotesize 307 (10.2) &\footnotesize 10.1 (1.6) [15.4] &\footnotesize 4.4\\
\footnotesize GL490 &\footnotesize 0.087 (0.087) &\footnotesize 9.7 (2.4) &\footnotesize 60 (11.8) &\footnotesize 6.2 (4.9) [14.8] &\footnotesize 9.5-13.2\\
\footnotesize GL2591 &\footnotesize 0.097 (0.055) &\footnotesize 11.9 (2.4) &\footnotesize 131 (3.11) &\footnotesize 11.0 (1.3) [16.4] &\footnotesize 15.7-27\\
\footnotesize NGC2071 &\footnotesize 0.051 (0.044) &\footnotesize 4.2 (1.5) &\footnotesize 13.2 (3.54) &\footnotesize 3.1 (2.4) [7.3] &\footnotesize 7.7\\
\footnotesize W3(Main) &\footnotesize 0.39 (0.20) &\footnotesize 29 (7.5) &\footnotesize 204 (16.2) &\footnotesize 7.1 (2.2) [12.2] &\footnotesize 18.4\\
\footnotesize Cep A &\footnotesize 0.11 (0.11) &\footnotesize 11.3 (3.0) &\footnotesize 156 (15.3) &\footnotesize 13.8 (5.1) [25.1] &\footnotesize 54\\
\footnotesize W28(A2) &\footnotesize 0.14 (0.14) &\footnotesize 10 (3.4) &\footnotesize 889 (82.7) &\footnotesize 88 (24) [149] &\footnotesize 8.9\\
\footnotesize NGC6334 I(N) &\footnotesize 0.22 (n/a) &\footnotesize 9.5 (n/a) &\footnotesize 312 (n/a) &\footnotesize 33 (n/a) [44] &\footnotesize 25\\
\hline
\end{tabular}
\end{center}
\footnotesize $^a$ Data from C93, except for NGC6334 I(N) (Megeath \& Tieftrunk 1999).\\
\footnotesize $^b$ C93 outflow data are divided into high (HV) and extremely high (EHV) velocity parts.\\
\footnotesize $^c$ Includes correction factor of 1.33 for atomic gas (TM02).
\end{table}

%cfm As far as I can determine, the proper abbreviation for Cepheus A
%is Cep A.
Excluding the cases of Cep A and NGC6334 I(N) (below), we find the
logarithmic mean value of $\eta$ is 10.4. The standard deviation is
about a factor of two, consistent with the observational
uncertainties. This gives a cluster formation time (eq.[\ref{tform}])
of $7.1\times 10^5 (M_{\rm cl,typ}/1000\:{\rm M_\odot})^{1/4}
\Sigma^{-3/4}\:{\rm yr}$, which is consistent with estimates of $\la
10^6\:{\rm yr}$ from the spread in pre-main-sequence ages in the Orion
Nebula Cluster (Palla \& Stahler 1999), which has $M_{\rm cl}\sim
4500\:{\rm M_\odot}$ (Hillenbrand \& Hartmann 1998). We note that the
effect of wind-wind interactions, which we have ignored in our
analysis, would shorten this estimate by the fraction of wind momentum
that is dissipated.

Cep A is the most extreme of the C93 protoclusters in terms of its
surface density. The estimates of virial mass and surface density are
made uncertain by the fact that 2D maps of the dense gas show moderate
elongation (Y. Shirley, private communication). Furthermore, the
outflow observations do not symmetrically probe the central region of
Cep A (M. Choi, private communication) and so the total
$\dot{p}_{w,\rm tot}$ may be underestimated, thus artificially raising
$\eta$. For these reasons we have excluded Cep A from our statistical
sample. NGC6334~I(N)'s outflow was observed by Megeath \& Tieftrunk
(1999) and the outflow parameters determined using different
techniques to those of C93.  This source is thought to be in an
extremely early stage of evolution; for example the observed outflow
extends over only a small fraction of the clump, which has a
relatively low luminosity. Again we exclude this source from the
statistical sample, but consider it in the context of an example of
the earliest stage of cluster formation.

A common observational diagram used in the study of outflows from
protostars and protoclusters plots $\dot{p}_{w,\rm tot}$ versus
$L_{\rm bol}$. We consider a simple description of the evolution of
protoclusters %forming at constant star formation rate (eq.[\ref{sfr3}]) 
in this diagram. Assuming a constant star formation rate (eq.[\ref{sfr3}])
implies that, averaged over a population of clusters, $\dot{p}_{w,{\rm
tot}}$ is a constant over most of the evolution. 
%cfm I do not understand the last sentence.  Constant with respect
%to what?  Eq sfr3 has Mdot in it, not p_w anyway.
In the earliest stages the luminosity is small and dominated by that
resulting from accretion. 
%For a given $\dot{p}_{w,\rm tot}$, derived
%from equation (\ref{pdotmedian}), 
The minimum luminosity is $\sim G \overline{m}_*
\dot{M}_*/r_*(\overline{m}_*)$, where $\overline{m}_*\la
\overline{m}_{*f}$ is the mean
protostellar mass. We have $\dot{M}_*= \dot{p}_{w,\rm tot}/(\phi_w
\overline{v}_K)$, where $\overline{v}_K$ is the Keplerian velocity at
the stellar surface of an average protostar. 
%The precise value of
%$\overline{m}_*$ depends on the stage of evolution of the cluster, but
%after a short time $\sim t_{*f}$, should be $\la \overline{m}_{*f}$. 
The mean radius is about $3\:{\rm R_{\odot}}$
(Figure 1) and thus
\begin{equation}
\label{minL}
L_{\rm min}=4.3\times 10^{4}\left(\frac{0.5}{\phi_{w}}\right) \left(\frac{\overline{m}_*}
{0.3\:{\rm M_\odot}} \frac{3\:{\rm R_\odot}}{r_*(\overline{m}_*)}
\right)^{1/2} \left(\frac{\dot{p}_{w,\rm tot}}{\rm M_\odot 
km\:s^{-1}yr^{-1}}\right)\:{\rm L_\odot}.
\end{equation}
Observed protoclusters, selected for showing signs of massive star
formation, will tend to be over-luminous compared to equation
(\ref{minL}).  Those in the earliest stages of evolution should lie
systematically closer to the birthline.
%which is essentially independent of $\Sigma$.
%(minor variations are
%expected in the mean protostellar radius as $\Sigma$ and thus the
%typical accretion rate changes; Stahler 1988). This minimum
%luminosity defines the ``protocluster birthline''.

The maximum luminosity is approximately the post-accretion luminosity,
$L_{*f}$, for all but the smallest clusters (where accretion can make
a significant contribution). We have determined $L_{*f}(M_{*f})$
from Monte Carlo simulations (above). Relating 
%cfm
$M_{*f}=2 M_{\rm cl,typ}/3$ for a typical forming cluster (TM02), 
%cf
we then eliminate $M_{\rm cl,typ}$ in equation
(\ref{pdotmedian}). The resulting expression defines the ``cluster
birthline'' (the 3 cases correspond to values of $M_{*f}/{\rm
M_\odot}$ of $10^2 - 10^3$, $10^3-1.6\times 10^3$ and $>1.6\times
10^3$, respectively):
%cfm Can't you reduce this to just two cases??
\begin{equation}
L_{*f} = \cases{
3.5\times 10^{7} \phi_L^{-2.0} \Sigma^{-1.5}
\left(\frac{\dot{p}_{w,\rm tot}}
{\rm M_\odot km\:s^{-1} yr^{-1}}\right)^{2.0}{\rm L_\odot} 
& $6.9\times 10^3\la \frac{L_{*f}}{\rm L_\odot}\la 4.4\times 10^5$\cr
7.9\times 10^{7} \phi_L^{-2.4} \Sigma^{-1.8}
\left(\frac{\dot{p}_{w,\rm tot}}
{\rm M_\odot km\:s^{-1} yr^{-1}}\right)^{2.4}{\rm L_\odot} 
& $4.4\times 10^5\la \frac{L_{*f}}{\rm L_\odot}\la 1.0\times 10^6$\cr
1.1\times 10^{7} \phi_L^{-4/3} \Sigma^{-1} \left(\frac{\dot{p}_{w,\rm
tot}}
{\rm M_\odot km\:s^{-1} yr^{-1}}\right)^{4/3}{\rm L_\odot} 
& $1.0\times 10^6\la \frac{L_{*f}}{\rm L_\odot}$}
\label{maxL}
\end{equation}
where $\phi_L=(\phi_w/0.5)(10/\eta)$.
%cfm Note replacement of X with \phi_L and removal of f_* and
%rearrangement of 1.0 times 10^6

The protocluster and cluster birthlines are shown in Figure 3.  In
Figure~3a we also plot the results of Monte Carlo simulations of many
clusters for two cases: $M_{*f}=200, 1000\:{\rm M_\odot}$. For each
cluster, stars are drawn randomly from the IMF and stochastically in
time and then followed through our protostellar evolution model for
pressures corresponding to $\Sigma=1\:{\rm g\:cm^{-2}}$. The
central/outer lines show the median/68-percentile values of each
cluster distribution.  Points mark equal time intervals in the
evolution.  In Figure~3b we plot the observed protoclusters, giving
$\dot{p}_{w,\rm tot}$ a factor of four uncertainty.  Sources in an
early stage of formation (NGC2071 and NGC6334 I(N)) are close to the
protocluster birthline, while those in a late stage (W3(${\rm
H_2O}$)---as evidenced by its multiple \ion{H}{2} regions) are close
to or below the cluster birthline.

\begin{figure}  
\label{fig:pdotL4}
\plottwo{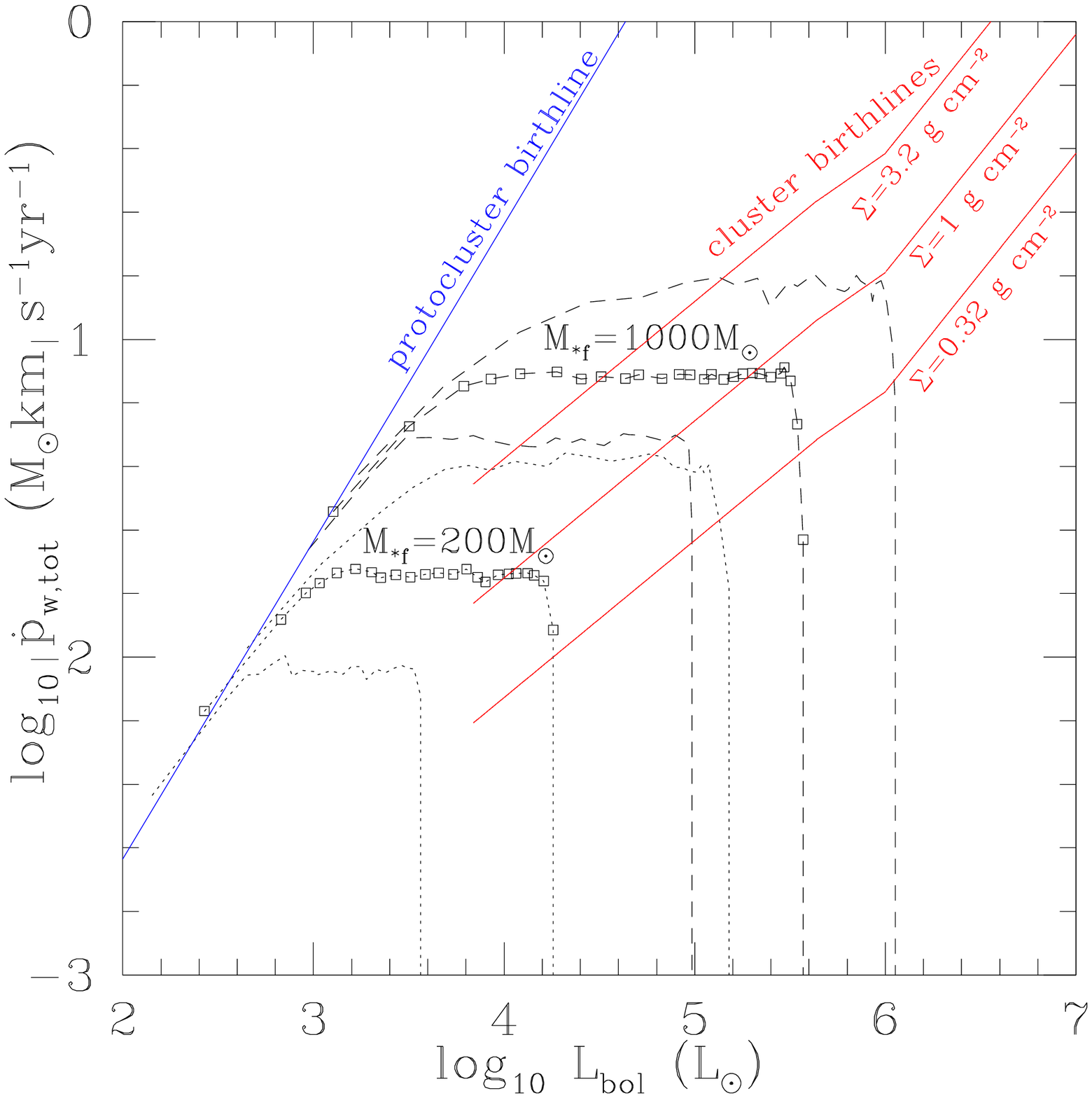}{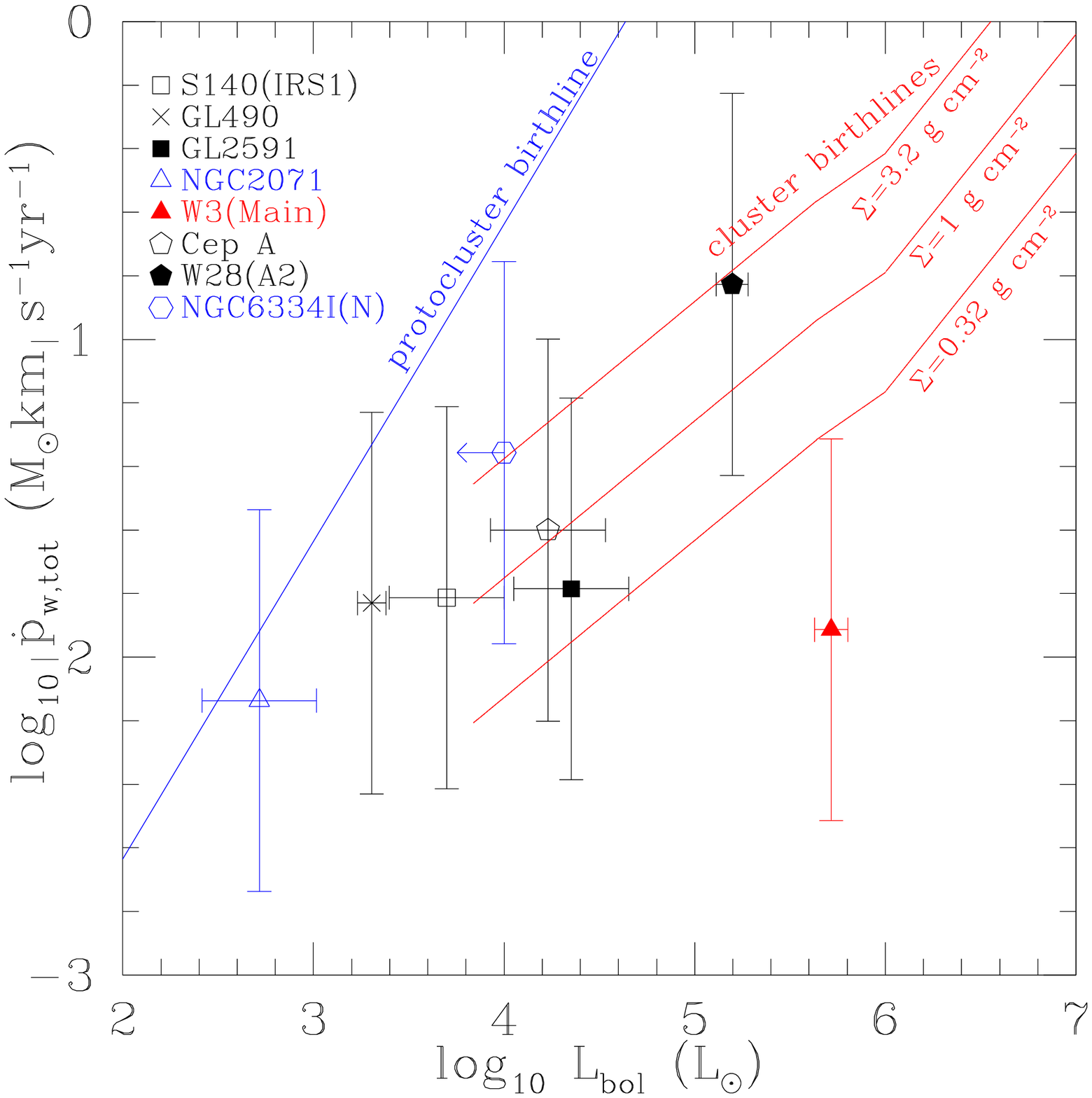}
\caption{$\dot{p}_{w,\rm tot}$ versus $L_{\rm bol}$. (a) Left: Theoretical birthlines and results of Monte Carlo simulations. (b) Right: Observed protoclusters.}
\end{figure}

\section{Conclusions}

We have presented a theoretical model for high-mass star formation
that determines the accretion rate in terms of the final and
instantaneous masses of the star, the ambient pressure surrounding the
gas core, and the polytropic index of pressure support. We related the
pressure due to self-gravity in the centers of typical Galactic
star-forming clumps to their surface density; $\overline{P}/k$
approaches $10^9\:{\rm K\:cm^{-3}}$ for typical values of
$\Sigma\simeq 1\:{\rm g\:cm^{-2}}$. These pressures allow a $100\:{\rm
M_\odot}$ star to form in $\sim10^5\:{\rm yr}$ with a final accretion
rate $\sim10^{-3}\:{\rm M_\odot\:yr^{-1}}$.  Modeling protostellar
evolution, we predicted the properties of several nearby massive
protostars. Under the hypothesis that the outflows from massive
protostars are driven magneto-centrifugally in the same manner as
those from low-mass protostars, we found that each decade of the IMF
contributes approximately equally to the total outflow momentum. This
helps to explain the generally poor overall collimation of outflows
from high-mass star-forming regions. We utilized these models and
observations of outflow intensities from several nearby protoclusters
%for star
%formation and outflow generation 
to estimate their star formation rates, finding that these clusters are typically
forming over at least several free-fall times. We have proposed
theoretical birthlines in the $\dot{p}_{w,\rm tot}$ versus $L_{\rm
bol}$ diagram to delimit the phase of cluster formation.

\acknowledgments We thank Henrik Beuther, Minho Choi, Neal Evans, Ralph Pudritz,
Yancy Shirley, Steve Stahler and Malcolm Walmsley for helpful
discussions. Our research is supported by NSF grants AST-9530480 and
AST-0098365,
%cfm I hope we have enough space for this:
by a NASA grant supporting the Center for Star Formation Studies,
and (for JCT) by a Spitzer-Cotsen fellowship from Princeton University.

\end{document}